\documentclass[letterpaper,twocolumn,10pt]{article}
\usepackage{usenix-2020-09}

\usepackage{tikz}
\usepackage{amsmath}
\usepackage{pgf-pie}  
\usepackage{array, multirow, graphicx, booktabs}
\usepackage{float}

\usepackage{pifont}
\usepackage{wasysym}
\usepackage{flushend}
\newcommand{\xmark}{\CIRCLE}%
\newcommand{\hmark}{\LEFTcircle}%

\usepackage{url}

\usepackage[inline]{enumitem}

\usepackage{tikz,colortbl}
\usetikzlibrary{calc}
\usepackage{zref-savepos}

\newcounter{NoTableEntry}
\renewcommand*{\theNoTableEntry}{NTE-\the\value{NoTableEntry}}

\newcommand*{\notableentry}{%
	\multicolumn{1}{@{}c@{}|}{%
		\stepcounter{NoTableEntry}%
		\vadjust pre{\zsavepos{\theNoTableEntry t}}% top
		\vadjust{\zsavepos{\theNoTableEntry b}}% bottom
		\zsavepos{\theNoTableEntry l}% left
		\hspace{0pt plus 1filll}%
		\zsavepos{\theNoTableEntry r}% right
		\tikz[overlay]{%
			\draw[black]
			let
			\n{llx}={\zposx{\theNoTableEntry l}sp-\zposx{\theNoTableEntry r}sp},
			\n{urx}={0},
			\n{lly}={\zposy{\theNoTableEntry b}sp-\zposy{\theNoTableEntry r}sp},
			\n{ury}={\zposy{\theNoTableEntry t}sp-\zposy{\theNoTableEntry r}sp}
			in
			(\n{llx}, \n{lly}) -- (\n{urx}, \n{ury})
			(\n{llx}, \n{ury}) -- (\n{urx}, \n{lly})
			;
		}% 
	}%
}

\begin{document}

%don't want date printed
\date{}

% make title bold and 14 pt font (Latex default is non-bold, 16 pt)
%\title{\Large \bf RPKI RRDP Threat Model from a Relying Party Perspective}
%\title{\Large \bf rpkiller -- Disrupting RPKI Relying Party software from within the tree}
\title{rpkiller: Threat Analysis from an RPKI Relying Party Perspective}

%for single author (just remove % characters)
\author{
	%Anonymous Authors
{\rm Koen van Hove, Jeroen van der Ham and Roland van Rijswijk-Deij}\\
Faculty of Electrical Engineering, Mathematics and Computer Science \\
University of Twente, the Netherlands
} % end author

\maketitle

% !TEX root = ../paper.tex

\begin{abstract}
	% ~150 words, 200 max
	The Resource Public Key Infrastructure (RPKI) aims to secure internet routing by creating an infrastructure where resource holders can make attestations about their resources. RPKI Certificate Authorities issue these attestations and publish them at Publication Points. Relying Party software retrieves and processes the RPKI-related data from all publication points, validates the data and makes it available to routers so they can make secure routing decisions. 

	In this work, we create a threat model for Relying Party software, where an attacker controls a Certificate Authority and Publication Point. We implement a prototype testbed to analyse how current Relying Party software implementations react to scenarios originating from that threat model. Our results show that all current Relying Party software was susceptible to at least one of the identified threats. In addition to this, we also identified threats stemming from choices made in the protocol itself. Taken together, these threats potentially allow an attacker to fully disrupt all RPKI Relying Party software on a global scale. We performed a Coordinated Vulnerability Disclosure to the implementers and have made our testbed software available for future studies.
\end{abstract}

% !TEX root = ../paper.tex

\section{Introduction}\label{sec:introduction}

The internet consists of \textit{inter}connected \textit{net}works managed by many different organisations. Data is exchanged within these networks, but also between these networks, e.g., data from Google to a residential customer at Comcast. Just like addressing physical mail, the data on the internet needs an address. This address is generally an Internet Protocol (IP) address, of which two versions are currently in wide use, namely IPv4 and IPv6. Much like physical mail, there are arrangements between operators to handle each other's data. The Border Gateway Protocol (BGP) is a protocol created to coordinate this. Every network, also called an Autonomous System (AS), sends 
\begin{enumerate*}[label=\arabic*)]
	\item who they are;
	\item who can be reached (indirectly) via them
\end{enumerate*}. They send this to the peers they are connected to. Every network has an Autonomous System Number (ASN). By creating a list of these ASNs, one can eventually discover how to reach all destinations. This string of numbers, called an AS path, allows every network to find a path to any other network \cite{rfc4271}.

The main issue facing BGP currently is that it is largely based on good faith. By default, anyone can send any BGP announcement, with their own ASN as the destination for \textit{any} prefix \cite{goldberg_2014}. This means someone else than the owner is now in control of a set of IP addresses, and can use them for their own purposes. These situations are not purely hypothetical - they have occurred in the past due to misconfiguration or malice \cite{strickx_2019, goodin_2018}. BGP forms an integral part of the worldwide inter-AS routing system, and is not easily replaced. While a secure version of the BGP protocol exists (called BGPsec~\cite{rfc8205}), this protocol has gained little traction and is not actively used by operators due to deployment challenges~\cite{goldberg_2014, chung2019}.

To overcome the security challenges in inter-domain routing, the Resource Public Key Infrastructure, or RPKI for short, was introduced from 2011 onward. RPKI is a public key infrastructure framework that aims to work together with BGP \cite{rfc6480}. RPKI tries to secure routing in the same way that WebPKI tries to secure the connection between your web browser and the web server, where a certificate attests that you are actually talking to the website you intended to visit. 

% Using the standard method of longest prefix matching, if one announces a specific prefix that is more specific than the other announced prefixes, it will be used (although an operator may decide to ignore too specific announcements, generally for IPv4 anything more specific than /24, and for IPv6 anything more specific than /48 \cite{huston_2020}), and traffic will be routed to your ASN. Alternatively, when announcing the same prefix length, most autonomous systems will use the shortest path, thus part of the traffic may be rerouted to your ASN if the path to your ASN turns out to be shorter than the original path. What happens is that an autonomous system hijacks an IP prefix this way.

RPKI currently solves that issue by providing attestation using aforementioned certificates. There are five generally agreed upon certificate trust anchors, namely the five Regional Internet Registries (RIRs) -- AfriNIC, APNIC, ARIN, LACNIC, and RIPE. These five RIRs all host Route Origin Authorizations (ROAs) for their customers directly on their own publication point, but RPKI can also be delegated, where customers run their own CA and host their own publication point. So-called Relying Party (RP) software retrieves the data from the publication points recursively, validates it and then makes it available to a router so it can make secure routing decisions. Deployment of RPKI is increasing rapidly~\cite{chung2019}, especially when it comes to delegated RPKI. In the context of the latter, the question that arises is whether a dishonest delegatee can abuse their power to disrupt Relying Party software, and potentially RPKI as a whole.

For that reason, in this paper we ask ourselves: \textit{How can a dishonest Publication Point or Certificate Authority disrupt the operations of Relying Party software?} To answer this question, we consider the following sub-questions:

\begin{enumerate}
	\item What known threats exist to the stack that the RPKI runs on?
	\item Are there any exploitable security vulnerabilities caused by omissions in the RFCs regarding implementation considerations?
	\item What aspects of RPKI can a Publication Point and/or Certificate Authority abuse to make Relying Party software malfunction?
\end{enumerate}

\noindent
To answer these questions, we:

\begin{enumerate}
	\item Develop a threat model for RPKI focussing on the abilities of a dishonest publication point and/or certificate authority;
	\item Create a proof-of-concept exploit framework and an overview of the behaviour of relying party software with regards to the threats we identified.
	%\item Give a recommendation on how to counter the aforementioned threats.
\end{enumerate}

\noindent
\textbf{Outline --}
The remainder of this paper is organised as follows: in Section~\ref{sec:background}, we provide background on the RPKI. In Section~\ref{sec:related-work} we discuss related work. In Section~\ref{sec:threat-model} we introduce our threat model. In Section~\ref{sec:tests} we introduce practical test cases based on the threat model and test those against eight open source Relying Party implementations. In Section~\ref{sec:discussion} we reflect on the results of these tests, and discuss possible mitigations. Lastly, in Section \ref{sec:vulnerability-disclosure} we discuss ethical considerations and describe the process we used to notify RP implementers of the vulnerabilities we identified. 
% !TEX root = ../paper.tex

\section{Background}\label{sec:background}
\subsection{BGP}\label{sec:bgp}
The Border Gateway Protocol (BGP) is a protocol that aims to exchange routing and reachability information for Autonomous Systems (AS). BGP generally comes in two varieties: interior (within an AS) and exterior (between ASes). We only concern ourselves with the latter -- an AS has full control over the routing within their network, and is thus out of scope.

\begin{figure}[t]
	\includegraphics[width=1\linewidth]{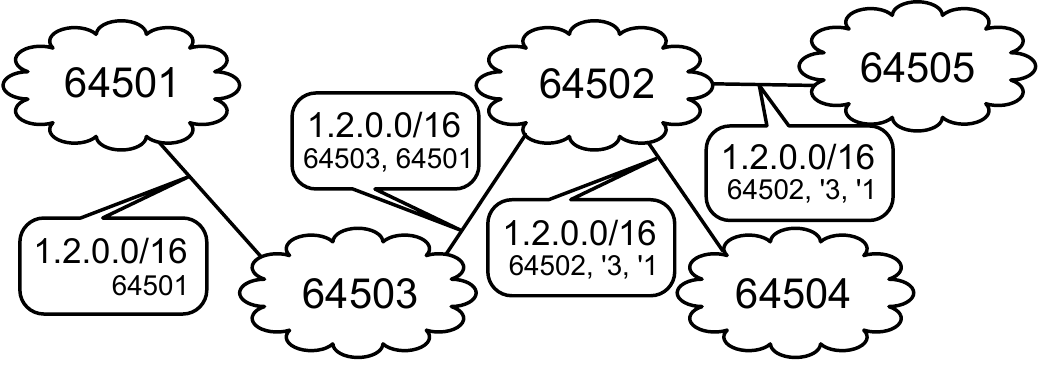}
	\centering
	\caption{An example of a network with five autonomous systems, {\tt 64501} to {\tt 64505}, with lines representing interconnections. An example of the BGP advertisement for {\tt 1.2.0.0/16} from {\tt 64501} is shown.}
	\label{fig:bgp}
\end{figure}

Let us explain how BGP works by means of example as shown in Figure~\ref{fig:bgp}. The nodes {\tt 64501} to {\tt 64505} represent our ASes, and the lines represent connections between them. Initially, the nodes do not know about each other. AS {\tt 64501} can {\it advertise} to {\tt 64503} that {\tt 64501} is the destination for {\tt 1.2.0.0/16}, giving an {\it AS path} of $\langle${\tt 64501}$\rangle$. AS {\tt 64503} learns that information, and advertises to {\tt 64502} that {\tt 1.2.0.0/16} can be reached via $\langle${\tt 64503, 64501}$\rangle$. Repeat this, and {\tt 64504} learns that a path to {\tt 1.2.0.0/16} is $\langle${\tt 64502, 64503, 64501}$\rangle$. This is depicted in Figure~\ref{fig:bgp}. If {\tt 64505} also advertises {\tt 1.2.0.0/16}, then {\tt 64504} will learn that $\langle${\tt 64502, 64505}$\rangle$ is also a route for that prefix. Generally a shorter route is preferred. In addition to this, BGP uses the principle of longest prefix matching, meaning if a more specific prefix were to be advertised, e.g., {\tt 64505} advertising {\tt 1.2.3.0/24}, then {\tt 64503} would generally prefer that over the more general announcement of {\tt 1.2.0.0/16} from {\tt 64501}. Note that in practice every AS has its own policy for routing, for example regarding the maximum prefix length. BGP merely makes claims about what \emph{can} be reached, not which path \emph{should} be used. An operator may thus pick a different path. 

% RvRD: commented out sentence add detail that isn't relevant for the point you're driving
% toward, which is what RPKI is for
%
%1.2.0.0/16 does not appear in the clouds in figure \ref{fig:bgp}. The reason for this is that the claim by both $64501$ and $64505$ is entirely based on good faith. 
BGP does not provide any means to verify any of the claims made by an AS. It is thus possible to claim to be the destination of a prefix owned by someone else, or to claim a different AS is the destination, or to claim that an AS has connections to other ASes which it does not have. 
%For example, $64505$ may advertise a path of $\langle 64505, 64501 \rangle$ for 203.0.113.0/24, even though that does not exist in real life and $64501$ is not the destination for that prefix.

\subsection{Increasing BGP security}\label{sec:bgp-security}

There have been many attempts at making BGP more secure, such as BGPsec \cite{rfc8205}. One of the main hindrances in most of them is that they require a fundamental change to the BGP protocol and (near) universal deployment, thus requiring a flag day. In contrast, RPKI is a separate infrastructure that can be implemented gradually, while already bringing security benefits. RPKI aims to enable creating attestations about general internet number resources, but in its currently form limits itself to so-called Route Origin Authorizations (ROAs)~\cite{rfc6482}. A ROA contains an ASN and one or more IP prefixes. Upon receipt of an announcement from a peer, a BGP speaker can look up whether there exists a ROA that covers this prefix. The result can be: 
\begin{enumerate*}[label=\arabic*)]
	\item valid - a ROA exists, and the origin ASN matches the ROA;
	\item invalid – a ROA exists, and the origin ASN does not match the ROA, or the prefix length does not match the length specified in the ROA;
	\item unknown (or ``not found'') – there is no ROA for this IP prefix
\end{enumerate*}~\cite{rfc6483}. This solves part of the good faith issue described above. It is important to note that a ROA only aims to protect the final destination for a BGP path (called the ``origin''), meaning that a valid destination does not guarantee an honest path~\cite{goldberg_2014}. Work is on the way to enable full path attestation through the RPKI with, for example, ASPA~\cite{ietf-sidrops-aspa-verification-07}. Additionally, it is to the recipient's full discretion what to do with the information from RPKI. They are free to ignore it entirely or partially.

%\subsection{RPKI infrastructure}\label{sec:rpki-infrastructure}
\begin{figure}[t]
	\includegraphics[width=1\linewidth]{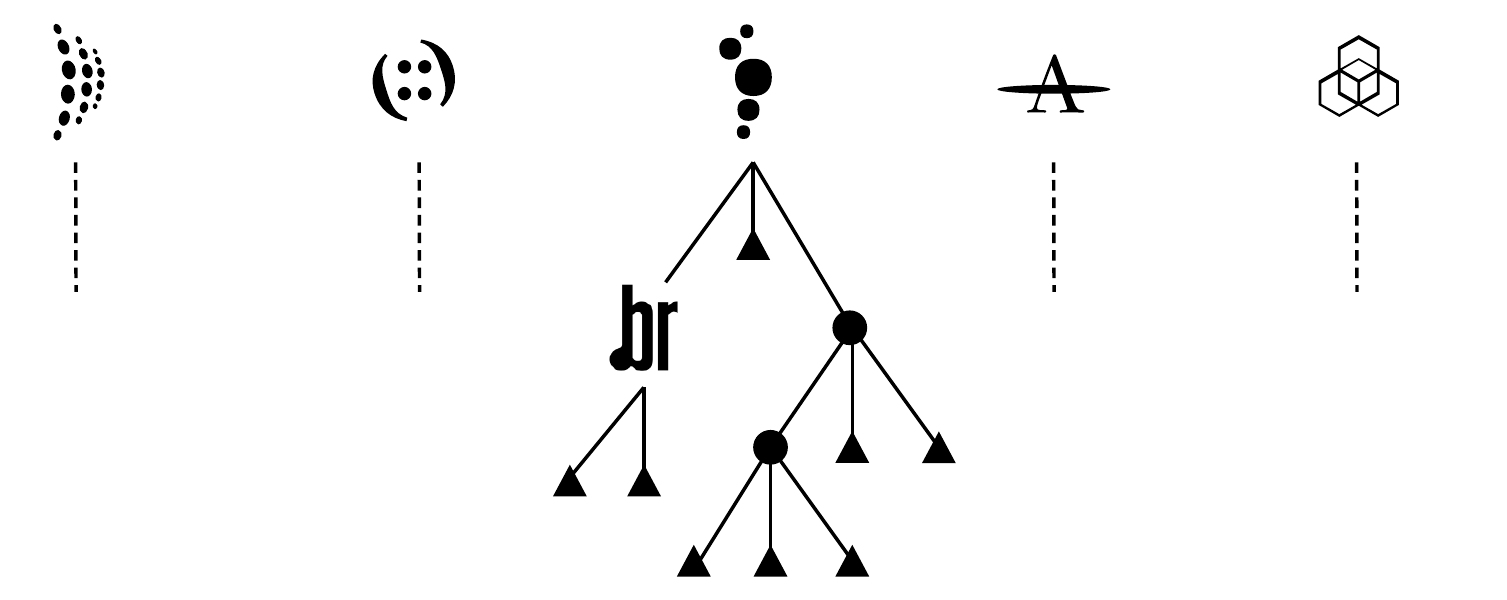}
	\centering
	\caption{Example of a tree showing the trust hierarchy in RPKI. The example tree originates from LACNIC. Every circle as well as the RIRs and the National Internet Registry (NIR) of Brazil (.br), represent a CA. Every triangle represents a ROA. Every subordinate CA can delegate (part of) their resources to further child CAs. The five trees of the five RIRs are in principle independent.}
	\label{fig:rpki}
\end{figure}

To determine whether a ROA is genuine, a standard PKI setup using X.509 certificates~\cite{rfc5280} is used. There are five generally agreed upon trust anchors in the RPKI, namely the five Regional Internet Registries (RIRs) – AfriNIC, APNIC, ARIN, LACNIC, and RIPE~\cite{rfc6480}. These five RIRs all host ROAs for their customers directly, but they may also delegate resources to subordinate certificate authorities, such as National Internet Registries (NIRs), or to organisations that prefer to manage their own ROAs. These subordinate certificate authorities can again create their own subordinate certificate authorities, and so on, and so forth. The RIRs also host a repository publication point (PP), which is the place where the ROAs, certificates, and other objects can be downloaded from. These publication points are referenced in the certificates~\cite{rfc6481}. Note that whilst normally a certificate authority and repository owner are managed by the same entity, this need not be the case. An example of a trust hierarchy tree can be seen in Figure~\ref{fig:rpki}. 

\begin{figure}[t]
	\begin{center}
		\includegraphics[width=0.5\columnwidth]{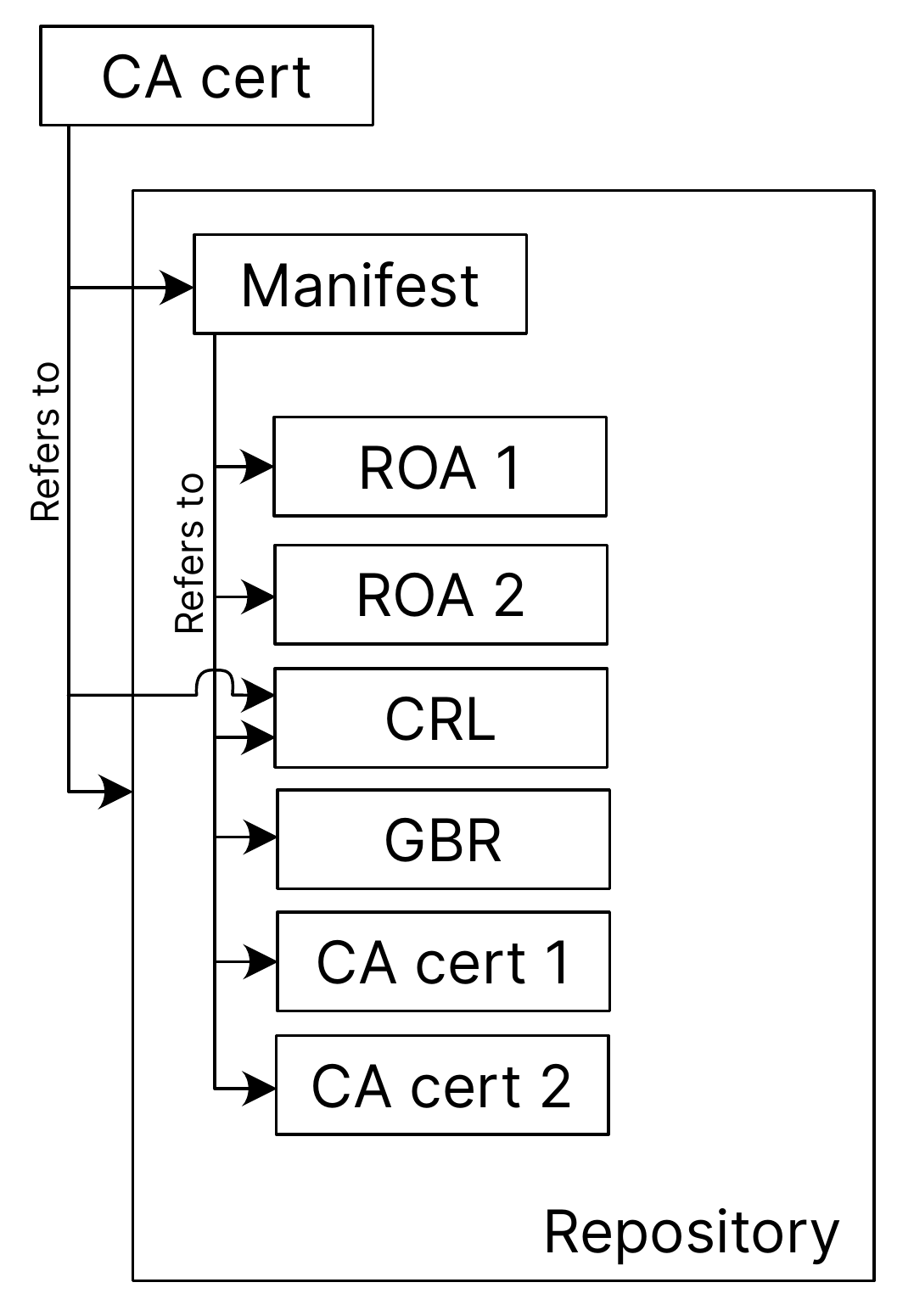}
	\end{center}
	\caption{\label{fig:repository}An example of a repository in RPKI. The CA cert contains an extension that states where the repository can be found, and what file inside the repository the manifest (MFT) is. The manifest then contains entries to the ROAs, Certificate Revocation List (CRL), Ghostbusters Record (GBR), and child CA certs. The objects in the repository (MFT, ROA, CRL, GBR, and CA certs) are CMS signed objects \cite{rfc5652} using a certificate signed by the CA.}
\end{figure}

In practice RPKI is deployed as follows: at the top of the tree are the five RIRs, each with their own root certificate and repositories. Those repositories, accessible via rsync or RRDP, a protocol specific to the RPKI based on HTTPS and XML, contain signed objects. Signed objects are currently mainly ROAs, although Ghostbusters records~\cite{rfc6493} have been added as well. These objects are signed with a certificate signed by the certificate authority. The repository also contains other certificates signed by the current certificate. Those certificates then point to their own repository, which is what gives it the tree structure. For every repository, there is a manifest file. A manifest file contains a list of expected signed objects and their hash - this can be used to check that the data from the repository is complete and correct~\cite{rfc6481}. An example of a repository can be seen in Figure~\ref{fig:repository}. 
% RvRD: is it important for readers to know the detail commented out here?
%Note that a single repository may host multiple manifests, and the CA certificates may refer to another manifest within the same repository.

Recently, more subordinate certificate authorities and repositories have appeared~\cite{kristoff_2020}. Relying party software, the software that collects all the ROAs and creates the lookup table, traverses these subordinate certificate authorities and their repositories. Previously, all repositories belonged to trusted parties, such as RIRs, NIRs, or a select few miscellaneous parties that were generally well-trusted. With the growth in delegated RPKI, this implicit faith in repository operators is no longer a given.

\subsection{CA and repository}\label{sec:ca-and-repository}
For the purpose of this work, it is important to differentiate between the notion of a repository and a CA:

\begin{description}
	\item[Repository] A repository, also called a publication point, is a place where the data from the RPKI can be found. A repository can be conceptualised as a ``folder'' with signed files (e.g., ROAs) in them.
	\item[Certificate Authority] A certificate authority (CA) is the holder of the private key that can sign the files.
\end{description}

In most cases the repository operator also the operates the certificate authority, but these can be two different entities. For example, an NIR can operate the repository publication point, whilst the actual signing is done by their customers directly. Additionally, in the case where RIRs host the ROAs for their customers directly, it is often the case that the RIR is both the repository and certificate authority, and the customer can only specify what objects the RIR should create.

%\subsection{Rsync}\label{sec:rsync}
% RvRD: it's a bit strange to talk about rsync explicitly, but not about RRDP;
% in the actual text you discuss both, and I think they point you're trying to
% make is that the paper will focus on RRDP since that is the go-to transport
% to use.
\subsection{RPKI communication protocols}\label{sec:rsync}

The RPKI supports two protocols for communication with publication points: rsync and RRDP. Initially, only rsync was used~\cite{rfc6480}, and support for rsync is, at the moment of writing, still required. All common relying party software, however, now also supports the RPKI Repository Delta Protocol (RRDP), as do nearly all of the repository publication points. There are several known issues associated with the use of rsync that have spurred the adoption of RRDP. These issues are not merely theoretical, but have been confirmed in practice~\cite{ietf-sidrops-prefer-rrdp-01}. We list important ones below:

\begin{enumerate}
	\item \emph{There is no formal specification of the rsync protocol.} Instead, rsync is specified by its implementation. All current relying party implementations use the reference implementation of rsync, with the exception of rpki-client, which uses an open client-side reimplementation of the early 2004 version of the rsync protocol (version 27) called OpenRsync~\cite{openrsync}. The use of rsync also adds a layer of difficulty for relying party software, as they need to interact with an external rsync process.
	\item \emph{Repositories that use rsync are plagued by non-atomic updates}~\cite{rfc8182, trenaman_2021}, i.e., if the content of a repository changes whilst a relying party is retrieving data, that attempt will fail as the manifest does not match the retrieved data.
	\item \emph{It is trivial to execute a successful denial-of-service attack on an rsync daemon} by having a desktop open several thousand connections to an rsync daemon from a /48 IPv6 subnet, different Tor exit nodes, several VPN endpoints, or a combination of all of these. This causes the rsync daemon to be overloaded, and blocks honest parties from retrieving the data from the repository over the rsync protocol. This violates the RPKI standard, RFC~6481, which states that ``The publication repository SHOULD be hosted on a highly available service and high-capacity publication platform.''~\cite{rfc6481}. Due to the reference implementation nature of rsync, we consider resolving this issue infeasible.
	\item \emph{Disrupting the operation of rsync clients is easy}, for example using a very large ``message of the day'', or serving a repository with millions of empty folders to cause inode exhaustion. This causes the relying party software to crash.
\end{enumerate}

Given these issues, as well as the current uptake of RRDP, we expect rsync to be fully replaced in practice by RRDP by the end of 2022. As a consequence, the main focus of this work will be RRDP, and we only consider rsync if the issue transcends the protocol used.

\section{Related work}\label{sec:related-work}

{\bf Threat models for IETF protocols} -- while RFCs typically include a section on security considerations, it is less common for IETF protocols to be analysed using (semi-)formal threat models. An example of an IETF protocol that, like BGP, is undergoing a security overhaul is the Domain Name System~\cite{rfc1034}. The introduction of DNSSEC~\cite{rfc4033} adds two security properties, authenticity and integrity, to the DNS. Designing DNSSEC took many years and a number of attempts at specifying a workable protocol. Over the course of the work on DNSSEC, Atkins and Austein~\cite{rfc3833} analysed threats to the DNS to help sharpen the discussions going into the design of the protocol. Other IETF work, by Bortzmeyer~\cite{rfc7626} and later by Wicinski~\cite{rfc9076} analyse privacy concerns in the DNS. Outside of the IETF some formal analysis of DNSSEC has been performed, e.g., by Bau et al.~\cite{bau_2010} who use the Mur$\phi$ model-checking tool to verify DNSSEC in a specific setting and uncover vulnerabilities that allow an attacker to insert a forged name under certain conditions.

As discussed in the previous section, for BGP itself, a secure variant has also been specified by the IETF. Its specification discuss potential threats to integrity and verifiability of data, but lacks a discussion of threats to availability~\cite{rfc8205}. In contrast, in this work, we specifically focus on availability threats to the RPKI.

A final example that stands out is TLS~1.3, which unlike previous versions of the Transport Layer Security protocol and its predecessor, SSL, does have well-described security models, and has undergone extensive formal verification~\cite{dowling_2021, rfc8446}. We note that these analyses also focus mainly on confidentiality, authenticity and integrity, but not on availability. Isolated cases of availability do, however, exist (e.g., \cite{cve-2019-6659}). An interesting example of what vulnerabilities can be uncovered in earlier versions of the TLS protocol -- that lack a well-described security model -- is given by De~Ruiter and Poll~\cite{ruiter2015}. Their work, which relies on automated protocol fuzzing, is even able to identify particular version of TLS implementations.

\vspace{0.5em}
\noindent
{\bf Studies of the RPKI} -- in the field of RPKI, the focus has mostly been on the external effects the infrastructure provides, instead of looking into the security of the infrastructure itself. For example, W\"{a}hlisch et al.~\cite{waehlisch_2015} look at the deployment of RPKI in the context of web hosting and which attacker models RPKI prevents. There has been research into the security of RPKI itself, such as the importance of consistency of objects~\cite{rfc6486}, the lifetime of objects and certificates~\cite{rfc6489}, and the use of the maxLength attribute~\cite{gilad_2017}. Work by Chung et al.~\cite{chung2019} studies growth in the deployment of ROAs. By comparing eight years of historical RPKI data to archived routing data from route collectors, they show that the quality of RPKI data (i.e. to what extent ROAs match BGP advertisements) has improved dramatically. Cooper et al.~\cite{cooper_2013} look at what power different entities within the RPKI ecosystem have, and what certain authorities could do to abuse that power. Shrishak et al.~\cite{shrishak_2020} also look at this, and come up with a solution that uses threshold signatures to limit the power of trust roots. Others have come up with solutions that involve the Blockchain~\cite{yan_2021} to counter the centralised nature of RPKI. Additionally, RFC~7132~\cite{rfc7132} mentions some security considerations regarding an adversarial CA, e.g., ``An attacker could create very deep subtrees with many ROAs per publication point, \ldots'', but does not assess the impact in detail, nor does it create a comprehensive analysis of threats to RPKI itself. 

% RvRD: speculation, I suggest leaving this out
%It seems implied that, although not stated explicitly, that this type of attack is from a purely theoretical nature due to the unlikelihood of a CA takeover, with an expectation that a CA can normally be trusted.

Regarding communication with RPKI repositories, there is little public security research into RRDP. The RRDP specification~\cite{rfc8182} does include security considerations, but those mostly concern integrity and tampering, with references to the recommendations for the use of TLS~\cite{rfc7525}, and some additional notes on how RRDP increases reliability. %RRDP is, however, built upon XML and HTTPS, and as we will show in this work, this has consequences for the security and availability of the protocol. 
% RvRD: it's nicer to end on a teaser for what we will do in our paper
% we consider it worth looking into whether the known vulnerabilities of those technologies affect RRDP as well.

% !TEX root = ../paper.tex

\section{Threat model}\label{sec:threat-model}

% RvRD: made explicit that we're looking to disrupt the availability of RPs, as
% this wasn't really being said anywhere.
In this section, we introduce our threat model. The goal of our model is to identify \emph{availability} threats to the RPKI, particularly threats that a dishonest CA can pose to the availability of relying party software. If an attacker succeeds in disrupting RP software, this may lead to BGP route hijacks succeeding, as a router is not provided with information on valid route origins. Another outcome may be denial of service, which can prevent new routes or updates from being accepted. If a ROA cannot be validated, parties relying on validation will then reject announcements based on stale information.

For this model, we assume that an attacker operates its own RPKI Certificate Authority and repository Publication Point, as part of an otherwise non-malicious RPKI tree and at the same level as other non-malicious organisations. Such an entity has the ability to sign new CAs and objects. We will first sketch the assumptions made about the RPKI in Section~\ref{sec:assumptions}, and then expand on the implications of those assumptions in Section~\ref{sec:implications}.

\subsection{Assumptions}\label{sec:assumptions}

Let us look at the implicit assumptions made about the RPKI, and whether these assumptions are guarded by technological measures, as part of the standards or not at all. We observe that three types of assumptions are generally made: 
\begin{enumerate*}[label=\arabic*)]
	\item about the size and depth of the RPKI tree hierarchy,
	\item on execution time of the validation process, and
	\item on the protocol stack RPKI communication protocols depend upon and the operating system relying party software runs on
\end{enumerate*}.
We discuss these three classes of assumptions below.

\begin{figure}[t]
	\includegraphics[width=\columnwidth]{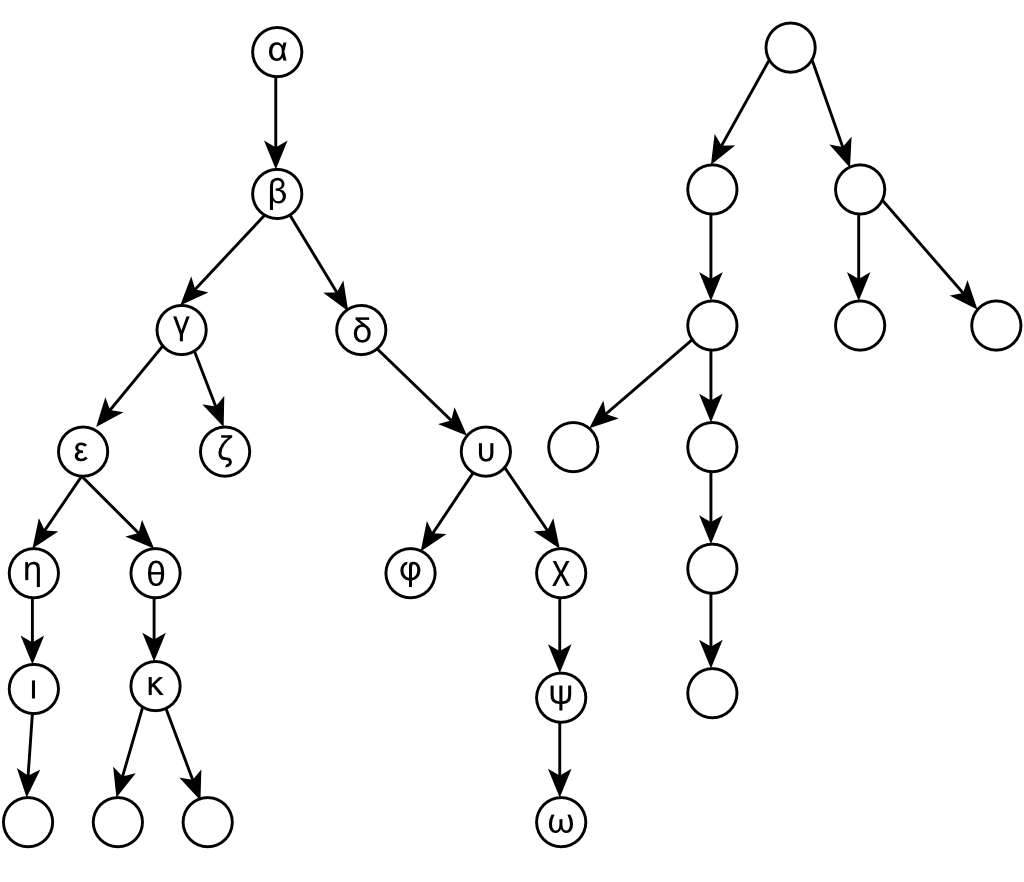}
	\centering
	\caption{An example of two RPKI trees. Each arrow signifies a parent-child relationship.}
	\label{fig:trees}
\end{figure}

\subsubsection{Tree size and depth}

When one considers the RPKI, the most common visualisation is a forest of trees as depicted in Figure~\ref{fig:trees}. The first assumption is thus that the RPKI is a forest of independent trees, where each tree stems from a Trust Anchor Locator (TAL), a reference to the root certificate preloaded into the relying party software. It must be possible to evaluate each tree in isolation, and trees must no contain loops. Furthermore, the RPKI validation process hinges on first collecting all information from all repositories before making routing decisions. For example, if a parent CA has a ROA for a {\tt /8}, and a child has a ROA for a subset of that {\tt /8} (e.g., a {\tt /16}) with a different ASN, and the child's repository is unavailable, then a route advertisement for that {\tt /16} might switch from valid to invalid (instead of unknown), which means routes become unavailable. If we take the tree in Figure~\ref{fig:trees}, and we suppose $\epsilon$ has a ROA for {\tt 1.0.0.0/8} with {\tt AS 1}, and $\eta$ has a ROA for {\tt 1.2.0.0/16} with {\tt AS 2}. If a BGP advertisement comes in for {\tt 1.2.0.0/16} with a path that ends with {\tt AS 2}, it is considered valid, as a known ROA for it exists. If we however do not consider the data from $\eta$, then that same advertisement would not match a known ROA, and instead, it would fall under the ROA from $\epsilon$ for {\tt 1.0.0.0/8} with {\tt AS 1}, which would make this BGP advertisement invalid (rather than unknown). % RvRD: should we explicitly mention this creates a potential availability DoS?

When we look at the nodes themselves, we make another set of assumptions. Let us look at node $\upsilon$ as shown in Figure~\ref{fig:trees}. We assume that the behaviour of node $\upsilon$ may affect itself, as well as its children $\phi$ to $\omega$, but that its behaviour may not influence the validity of its parents ($\alpha$,$\beta$ and $\delta$) nor its siblings ($\gamma$ and further down). Furthermore, a node must not require information from nodes other than its parents to function, thus $\upsilon$ must be able to work with only information from $\alpha$, $\beta$ and $\delta$. It is also assumed that all relying parties are served the same data by repositories, and thus have the same view on the RPKI.

% RvRD: I think what I wrote above is slightly clearer
%the view I have is the same as the view others have of the RPKI, and that no different data is returned based on the origin.

Summarising, this means that the assumption is made that the tree is finite and reasonably sized, and that that is the case for every relying party that fetches objects from repositories in the tree. 

\subsubsection{Time}

All of the assumptions about the tree size are there for one main reason: the certificates and objects in the RPKI all have an expiration time after which they are no longer valid. It is thus also assumed that the tree can be evaluated and processed in a reasonable amount of time, generally considered to be between 30 and 60 minutes~\cite{kristoff_2020}. If the process takes significantly longer, objects may expire before they can be evaluated. What is ``too long'' is undefined, but the shortest lifetime observed for objects in the wild is around 8~hours.

%Are the assumptions made about the RPKI justified? The independence of trees cannot be guaranteed -- a manifest can contain a certificate that has its parent in another tree at another TAL, or at another node in the same tree that is not its parent. This cannot lead to loops, as there can only be one Subject Information Access (SIA) and Authority Information Access (AIA) pair, which prevents cross-signing. This can only lead to loss of information for the node itself and its children, hence its impact is minimal. The other assumptions, like the tree being finite, reasonably-sized, and evaluable within reasonable time, are not guaranteed by the standards.

\subsubsection{Protocol stack and operating system}

As stated at the end of Section~\ref{sec:rsync}, when looking at communication protocols in the RPKI, we focus mostly on RRDP. The fact that RRDP is based on XML and HTTPS means that the threats that apply to that stack, namely XML, HTTPS, and TCP (or UDP in the case of HTTP/3~\cite{ietf-quic-http-34}) apply to RRDP as well. RRDP applies some extra restrictions, such as requiring the use of US-ASCII as character set, and forbidding the verification of the WebPKI certificate~\cite{rfc8182} on the TLS or QUIC layer. The signed objects themselves are CMS-encoded X.509 objects~\cite{rfc5652}, which means the threats to signed objects, certificates, and ASN.1 validation apply here as well. The only file in a repository that is not a signed object is a certificate to another repository in the tree. 

Lastly, the relying party software typically runs on a server, which most likely runs a UNIX-based operating system. This means that OS threats, such as inode exhaustion, running out of disk space, memory, or CPU cycles and other availability threats to the OS apply here as well.

\subsection{Implications}\label{sec:implications}

% RvRD: I reordered the implications in a (what I think is a) somewhat more
% logical order, starting with transport protocols, then what is transported,
% followed by RPKI structure and ending with the operating system.

Given the assumptions discussed in the previous section, we now discuss the implications and build our threat model to analyse threats by a dishonest CA to the availability of relying party software. We break down the threats in the sections below.

\subsubsection{HTTPS}\label{sec:https}

As stated before, modern RPKI implementations chiefly use RRDP to fetch objects from repository publication points. RRDP runs over HTTPS and thus potentially imports security issues from this protocol. In particular, we consider the following to be threats in the context of RPKI availability:

\begin{itemize}
	\item \emph{Decompression bombs} -- if the client supports compression such as GZIP, the server may serve a small file that is orders of magnitudes larger when unpacked~\cite{canet_2017};
	\item \emph{Redirect loops} -- returning a 301 HTTP status code with a location header with a location that either loops, or keeps forwarding the client to another location~\cite{owasp_rest};
	\item \emph{Forcing long retries} -- returning a 429 HTTP status code with a retry-after header with an unreasonably high value, such as a year or century~\cite{owasp_rest}.
\end{itemize}

\subsubsection{TCP}\label{sec:tcp}

Until HTTP/3 over QUIC becomes the norm, HTTPS relies on TCP as the underlying transport protocol. This means that RRDP also imports attacks on TCP. In the context of availability, we are most concerned about attacks that attempt to stall connections, such as a Slowloris attack \cite{cloudflare_slowloris}, or simulating a very slow unreliable connection.

\subsubsection{XML}\label{sec:xml}

RRDP uses XML as a message format. This makes it susceptible to attacks that exploit properties of XML~\cite{owasp_xml}. In the context of this work we consider:

\begin{itemize}
	\item \emph{Coercive parsing}, where an extreme (possibly endless) depth file is parsed;
	\item \emph{Invalid structures}, where the XML parser may trip up due to the unconventional nature of the file; 
	\item \emph{Very large payloads}, where the XML parser may load everything into memory, or temporarily store them on disk, causing resource exhaustion;
	\item \emph{Use of external entities}, thereby requesting resources the client may not be able to access, and possibly forwarding clients to a remote server;
	\item \emph{Infinite entity expansion}, by specifying recursive entities, or referential entities, an XML parser may run out of space or memory;
	\item \emph{Using external references} in XML to let the client DDoS another entity
\end{itemize}

\subsubsection{X.509}\label{sec:x509}

RPKI depends on X.509 at multiple levels (transport and content). X.509 is a complex standard, and the RPKI uses only a subset of its features. However, many relying parties use a standard library for X.509 that supports far more than RPKI requires. This makes it vulnerable to attacks such as specially crafted private keys \cite{cve-2018-1000613}. Additionally, given the binary nature of this format, unexpected data may cause the application to misbehave \cite{cve-2016-6308}. In the context of this work, we did not consider these potential vulnerabilities.
% RvRD: please check the statement I added at the end.

\subsubsection{ASN.1}\label{sec:asn1}

RPKI not only makes use of X.509, but also CMS, both of which are specified using ASN.1 and thus require parsers for objects that follow ASN.1 encodings such as the Basic and Distinguished Encoding Rules (BER and DER respectively). In the past, several major implementations, such as OpenSSL~\cite{cve-2016-2108} and BouncyCastle~\cite{cve-2019-17359}, had vulnerabilities related to parsing these encodings. Additionally, RPKI defines some uncommon modules with extra constraints, such as decoding an IP address prefix as {\tt BITSTRING}, which may not be adequately checked, and where strange values may cause undefined behaviour.

\subsubsection{Trees}\label{sec:trees}

RPKI repositories are assumed to be structured like a finite tree, where the application goes past every repository. This is similar to directory traversal, and several exploits exist here as well. A repository can, for example, introduce loops, or create an endless depth chain of certificates, thereby causing the relying party to endlessly fetch data. One can also introduce very many children, and thus go wide instead of deep. One can also combine these two approaches.

\subsubsection{Operating system}\label{sec:operating-system}

Finally, an attacker can attempt to target the underlying file system. Many Linux systems, for example, use {\tt ext4} as filesystem, which has a limited amount of inodes (pointers to entries) available. An attacker could serve a repository with many objects aiming to exhaust the available inodes. Alternatively, an attacker can try to serve extremely large files, e.g., a several gigabyte large Ghostbusters record to exhaust disk space~\cite{djordjevic_2011}.

% !TEX root = ../paper.tex
\renewcommand{\thesubsubsection}{\thesubsection.\Alph{subsubsection}}

\begin{figure}[t]
	\centering
	\includegraphics[width=\columnwidth]{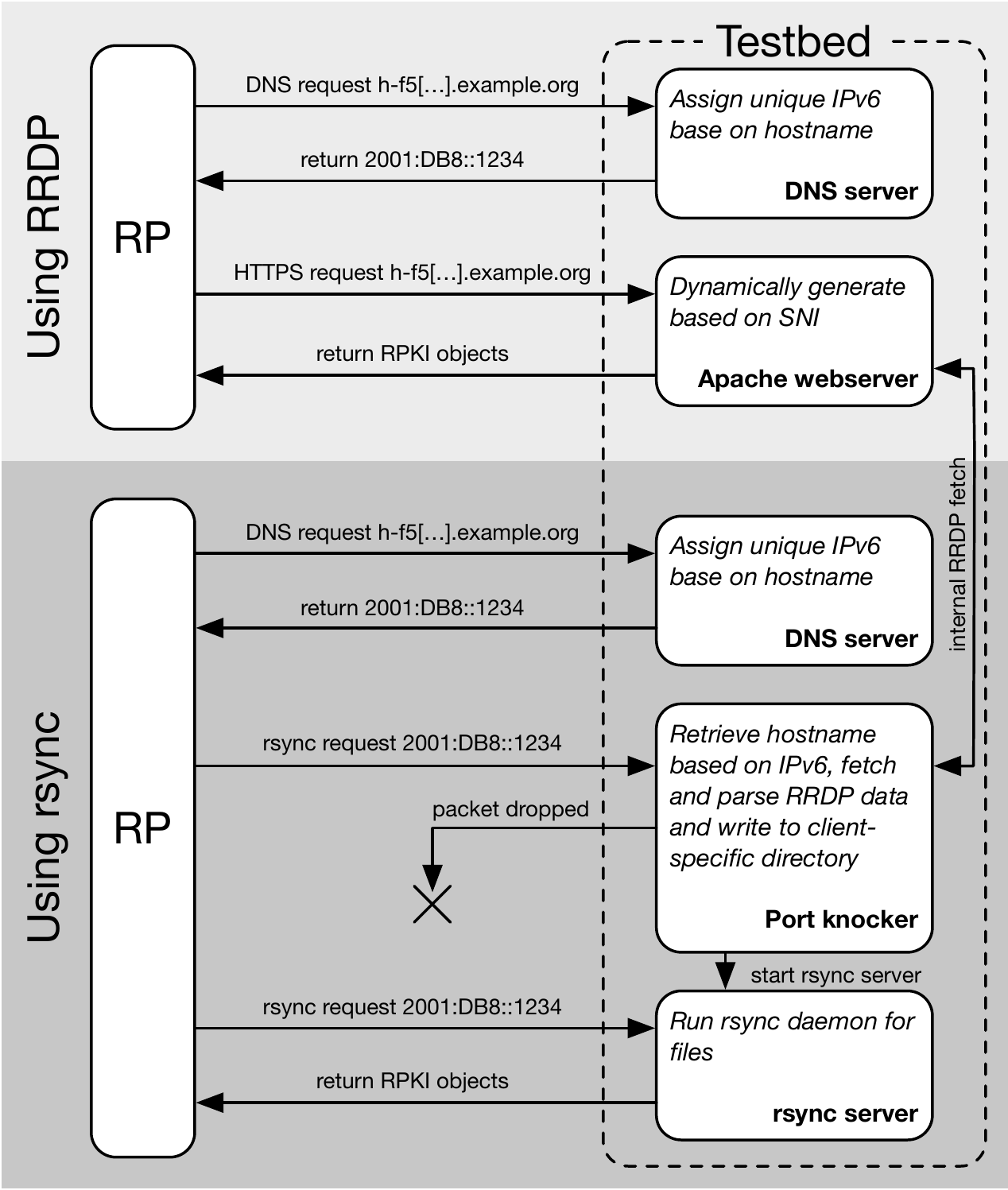}
	\caption{A graphical overview of how the testbed works, both for RRDP and rsync. For RRDP, the Server Name Indication (SNI)~\cite{rfc6066} extension is used, which contains the hostname that was requested. As this extension (or similar) is absent for rsync, the IP address used in the request is converted back to the corresponding hostname, and the testbed itself emulates the RRDP request. It then converts the RRDP output to files (if possible), and starts the rsync daemon. The original rsync request packet is dropped, but relying party (RP) software retries after a timeout.}
	\label{fig:testbed}
\end{figure}

\section{Evaluating threats in practice}\label{sec:tests}
% Evaluating threats in practice
In the analysis in Section~\ref{sec:threat-model}, we identified a set of potential vulnerabilities. To test if current relying party implementations are susceptible to these vulnerabilities, we designed and implemented a testbed that instantiates these vulnerabilities. In this section, we briefly introduce our testbed and discuss the tests we labelled A through O. We note that, in order not to disrupt any relying parties in production, our testbed functions independent from the existing RPKI hierarchy (discussed in more detail in Section~\ref{sec:vulnerability-disclosure}).

%The tests are labelled A to O, ordered chronologically.

\subsection{Testbed}\label{sec:testbed}

Our testbed creates custom repositories and CAs, and a separate trust anchor (TAL) is available for each test from A through O. The server creates the files for each instance on demand using a randomly generated UUID~\cite{rfc4122}. This prevents two test instances from interfering with each other, and also avoids the results from being influenced by caching. The resulting URI looks like \url{h-f5654a8f-6d17-4b62-9bb8-d5e11b8c08b2.example.org}, where the first letter of the hostname indicates the test (in this case ``H''), and the part after the first hyphen is the identifier. This host is assigned its own unique IPv6 address, and serves all requests for this test instance of test H over both RRDP and rsync. An overview of the setup can be found in Figure~\ref{fig:testbed}.

All relying party software implementations were installed on separate virtual machines, with standard configurations according to the instruction manual provided by the implementation. The version numbers of the RP implementations and operating systems of the VMs are listed in Table~\ref{fig:specs}. 

% !TEX root = ../paper.tex

\begin{table}[t]
	\centering
	\resizebox{\columnwidth}{!}{
	\begin{tabular}{|l|l|l|}
		\cline{2-3}
		\multicolumn{1}{l|}{} & \textbf{Version} & \textbf{Operating System} \\ \hline
		\textbf{Routinator}  & 0.10.1               & Ubuntu 18.04 LTS \\ \hline
		\textbf{Validator 3} & 3.2.2021.04.07.12.55 & Ubuntu 18.04 LTS \\ \hline
		\textbf{OctoRPKI}    & 1.3.0                & Ubuntu 18.04 LTS \\ \hline
		\textbf{Fort}        & 1.5.1                & Ubuntu 18.04 LTS \\ \hline
		\textbf{RPKI-Prover} & 0.1.0                & Ubuntu 18.04 LTS \\ \hline
		\textbf{rpstir2}     & master-7394f73       & Ubuntu 18.04 LTS \\ \hline
		\textbf{rpki-client} & 7.3                  & OpenBSD 7.0      \\ \hline
		\textbf{rcynic}      & 1.0.1544679302       & Ubuntu 16.04 LTS \\ \hline
	\end{tabular}
	} % resizebox
	\caption{Version information for the RP implementations that were tested.}
	\label{fig:specs} 
\end{table}

%\begin{table}[t]
%	\begin{center}
%		\begin{tabular}{|l|l|l|l|l|l|}
%			\cline{2-6}
%			\multicolumn{1}{l|}{} & \textbf{Version} & \textbf{Operating System} & \textbf{CPU}        & \textbf{RAM}              & \textbf{Disk}  \\ \hline
%			\textbf{Routinator}  & 0.10.1               & Ubuntu 18.04 LTS & 1× 2.3 GHz & 2 GB + 2 GB swap & 20 GB \\ \hline
%			\textbf{Validator 3} & 3.2.2021.04.07.12.55 & Ubuntu 18.04 LTS & 1× 2.3 GHz & 2 GB             & 20 GB \\ \hline
%			\textbf{OctoRPKI}    & 1.3.0                & Ubuntu 18.04 LTS & 1× 2.3 GHz & 2 GB + 2 GB swap & 20 GB \\ \hline
%			\textbf{Fort}        & 1.5.1                & Ubuntu 18.04 LTS & 1× 2.3 GHz & 2 GB             & 20 GB \\ \hline
%			\textbf{RPKI-Prover} & 0.1.0                & Ubuntu 18.04 LTS & 1× 2.3 GHz & 2 GB + 2 GB swap & 20 GB \\ \hline
%			\textbf{rpstir2}     & master-7394f73              & Ubuntu 18.04 LTS & 1× 2.3 GHz & 2 GB + 2 GB swap & 20 GB \\ \hline
%			\textbf{rpki-client} & 7.3                  & OpenBSD 7.0      & 1× 2.3 GHz & 2 GB             & 20 GB \\ \hline
%			\textbf{rcynic}      & 1.0.1544679302       & Ubuntu 16.04 LTS & 1× 2.3 GHz & 2 GB             & 50 GB \\ \hline
%		\end{tabular}
%	\end{center}
%	\caption{\label{fig:specs} Version information for the RP implementations that were tested. }
%\end{table}

\begin{figure}[b]
	\includegraphics[width=0.6\linewidth]{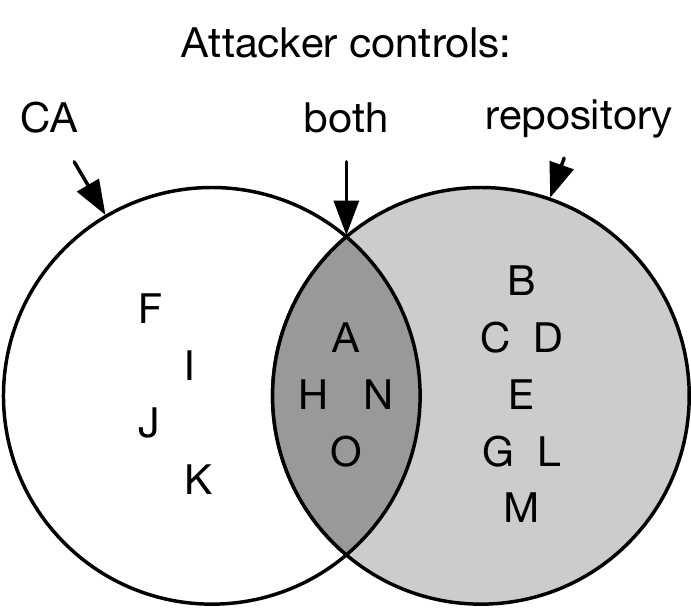}
	\centering
	\caption{Attacker control required for specific attacks.}
	\label{fig:venn}
\end{figure}

\subsection{Tests}

We will now discuss the tests we performed based on the identified threats. For clarity, the discussion of each test case is immediately followed by the results of the tests. As discussed earlier in the paper, the RPKI distinguishes between the roles of Certificate Authority and repository Publication Point operator. For the tests we describe, it is often not necessary for an attacker to have both roles. Figure~\ref{fig:venn} shows a Venn diagram depicting which role(s) an attacker needs to have in order to execute the attacks for each of our test cases.

Table~\ref{fig:test-table} summarises the results of the tests per relying party implementation. The details for each test are discussed in the sections that follow below. Note that the RIPE NCC RPKI Validator 3 and rcynic were not fully tested. In case of RIPE NCC RPKI Validator 3, the project was discontinued in July~2021~\cite{ripe-validator-end}, in case of rcynic, the last code commit at the time of writing of this paper occurred over 6~years ago~\cite{rcynic-repo}.

%The success of our tests is shown in table . Relying party software is constantly being updated; this research is thus not meant as a definitive analysis of current relying party software, but rather as an overview and threat model of dishonest RPKI repositories and/or certificate authorities, with the aim of hardening current and future relying party software. It thus might be that threats mentioned in this research do not impact current relying party software, but might impact future relying party software that adheres to the RFCs.

%The level of access an attacker would need to successfully execute the attack is shown in figure \ref{fig:venn}.

% !TEX root = ../paper.tex

\setlength\tabcolsep{1.8pt}
\begin{table}[t!]
	\begin{center}
		\begin{tabular}{|l|l|l|l|l|l|l|l|l|l|l|l|l|l|l|l|}
			\hline
			~ & \textbf{A} & \textbf{B} & \textbf{C} & \textbf{D} & \textbf{E} & \textbf{F} & \textbf{G} & \textbf{H} & \textbf{I} & \textbf{J} & \textbf{K} & \textbf{L} & \textbf{M} & \textbf{N} & \textbf{O} \\ \hline
			\textbf{Routinator} & \hmark & ~ & ~ & \xmark & \xmark & ~ & ~ & \xmark & ~ & \xmark & \xmark & ~ & ~ & \xmark & ~ \\ \hline
			\textbf{Validator 3} & \xmark & \notableentry & \notableentry & \notableentry & \notableentry & \notableentry & \notableentry & \xmark & \notableentry & \notableentry & \notableentry & \notableentry & \notableentry & \notableentry & \notableentry \\ \hline
			\textbf{OctoRPKI} & \xmark & ~ & ~ & \xmark & \xmark & \xmark & ~ & \xmark & \xmark & \xmark & \xmark & \xmark & ~ & \xmark & \xmark \\ \hline
			\textbf{Fort} & ~ & ~ & ~ & ~ & \xmark & ~ & ~ & \xmark & ~ & \xmark & \xmark & \xmark & ~ & ~ & \xmark \\ \hline
			\textbf{RPKI-Prover} & \xmark & ~ & ~ & ~ & ~ & ~ & \xmark & \xmark & ~ & \xmark & \xmark & ~ & ~ & ~ & ~ \\ \hline
			\textbf{rpstir2} & \xmark & ~ & ~ & \xmark & \xmark & \xmark & ~ & \xmark & \xmark & \xmark & \xmark & \xmark & ~ & \xmark & \xmark \\ \hline
			\textbf{rpki-client} & ~ & ~ & ~ & ~ & \xmark & ~ & \hmark & \xmark & ~ & \xmark & \xmark & \xmark & ~ & ~ & ~ \\ \hline
			\textbf{rcynic} & \notableentry & \notableentry & \notableentry & \xmark & \notableentry & \notableentry & \notableentry & \xmark & \notableentry & \notableentry & \notableentry & \notableentry & \notableentry & \notableentry & \notableentry \\ \hline
		\end{tabular}
	\end{center}
	\caption{The tests executed during our research in the then-recent relying party software. \xmark~means an implementation was vulnerable. \hmark~signifies a vulnerability that was fixed in a version that was released during our study. The crossed-out cells for Validator 3 and rcynic were not tested.}
	\label{fig:test-table} 
\end{table}
\setlength\tabcolsep{6pt}

\subsubsection{Infinite repository chain}

We create a chain of repositories, where the repository root has a certificate for a child, and when that child is visited, a certificate for this child is generated. This is repeated {\it ad infinitum}, creating a never-ending repository chain. This test case tries to attack the assumption that the tree is finite, and forces a client that does not impose limits to keep retrieving new data forever. 

\vspace{0.5em}
\noindent
{\bf Result} --
Most tested implementations keep retrieving data endlessly. Initial tests showed only Fort and rpki-client limit the maximum depth of a repository. A later version of Routinator, that was released while our study was still ongoing, also added a maximum depth.

\subsubsection{429 response header}
The {\tt 429} HTTP status code was added in RFC~6585~\cite{rfc6585}. It specifies that a user has sent too many requests, and should apply rate limiting. To provide a hint to the user when they can try again, a ``Retry-After'' may be included in the response, specifying how many seconds a user should wait before retrying. RFC~8182~\cite{rfc8182} does not specify whether this HTTP status code can appear in RRDP, and as relying party software tends to use existing HTTP libraries that may abstract this away, we wanted to find out what would happen if this value was set to an unreasonably long duration, for example a day. Older implementations sometimes instead use the 503 status code with Retry-After header.

\vspace{0.5em}
\noindent
{\bf Result} --
All tested implementations do not support 429 rate limiting, and neither do the libraries that they use.

\subsubsection{Endless 302 response}
The {\tt 3xx} range of HTTP status codes indicate that the resource can be found at another location~\cite{rfc1945}. These status codes can be chained, i.e., location (1) can link to (2) which links to (3). Much like test A, this can be done {\it ad infinitum}, thereby endlessly redirecting without technically looping. Relying party software is not required to support {\tt 3xx} status codes, nor is limiting the maximum amount of redirects required. In most cases the default value from the HTTP library is used.

\vspace{0.5em}
\noindent
{\bf Result} --
All tested implementations either do not support {\tt 3xx} redirects, or if they do, they limit it to a small amount, generally below 10.

\subsubsection{GZIP bomb}

HTTP allows for compression to improve transfer speed~\cite{rfc2616}. This means that the amount of data transferred may be less than the decompressed size on disk. Gzip is such a compression algorithm, and it is possible to create a file that is orders of magnitudes larger uncompressed than it is compressed, thereby exhausting the resources of the system the relying party software runs on. Support for compression is generally explicitly enabled in the HTTP libraries used by relying party implementations. 

\vspace{0.5em}
\noindent
{\bf Result} --
Routinator, OctoRPKI, rpstir2, and rcynic crash due to running out of memory, whereas Fort, rpki-client, and RPKI-Prover reject the repository.

\subsubsection{Open connection}

Repository publication points provide data at a certain bandwidth -- a minimum bandwidth is required to adhere to the reasonable time requirement. It is possible to break this bandwidth assumption by severely restricting transfer speed, for example to 3~bytes per second. This potentially makes the process of retrieving data run for several weeks if relying parties wait for completion. Note that data is constantly transmitted, just at a very slow rate, to avoid middleboxes cutting off the connection.

\vspace{0.5em}
\noindent
{\bf Result} --
All tested implementations apart from RPKI-Prover keep waiting forever. RPKI-Prover imposes a maximum transfer duration, and continues with the next repository once this timer expires.

\subsubsection{Broken ROA}

Whenever relying party software encounters a ROA --- a file with a {\tt .roa} extension --- it expects this file to have the structure that belongs to a ROA. However, it is possible to encode any data, and give it a {\tt .roa} extension. For this test case, the ROA merely consists of an encoded ASCII {\tt NUL} character. A lack of proper input validation may cause relying party software to malfunction or crash when they encounter this broken ROA.

\vspace{0.5em}
\noindent
{\bf Result} --
Only OctoRPKI and rpstir2 crash when encountering the broken ROA, whereas all other tested implementations show a warning and move on.

\subsubsection{Billion laughs attack}

The Billion Laughs attack is an XML expansion attack, where an entity ``lol1'' is defined as ``lol'', and ``lol2'' as ten times ``lol1''. This is done ten times to create a billion entities~\cite{owasp_xml}. The textual form is small, but XML parsers that try to handle this in memory will likely run out of memory and crash. 

\vspace{0.5em}
\noindent
{\bf Result} --
Only RPKI-Prover and rpki-client $\leq7.2$ crash when encountering the billion laughs attack. All other tested implementations treat this as a broken repository, and retry it over rsync.

\subsubsection{Exponential expansion}
\label{sec:testh}

Similar to test A, we create a chain of repositories. Instead of having one child, we have 10 children at each level. This means that the amount of CAs and repositories that must be visited becomes $\sum_{i=0}^{d} w^{i}$, where $w$ is the amount of children and $d$ is the depth. Even at a modest depth of~8, and a width of~10, this already becomes 11,111,111~repositories that must be visited. Even at one repository per second, this will take months, which breaks the expectation that the tree traversal can be done in a reasonable time.

\vspace{0.5em}
\noindent
{\bf Result} --
All implementations keep retrieving repositories. The order in which the repositories are retrieved does differ. We observed that most use either a depth-first search or breadth-first search approach, and that some will retrieve repositories in concurrent fashion. 

\subsubsection{Impossible ROA}

This test is similar to test F, but instead of a broken ROA, we present a syntactically correct ROA that is semantically broken. We do this by undermining the assumptions made in the structure of the ROA, namely, that an IPv4 address will have a prefix with at most 32 bits and an IPv6 address has a prefix with at most 128 bits. This is similar to the out-of-bounds vulnerability as described in CVE-2021-3761~\cite{cve-2021-3761}, and we expect a similar impact: either the validator crashes, or the output becomes nonsensical, causing the RPKI to Router (RTR) protocol, which facilitates communication between the relying party software and the router, to terminate.

\vspace{0.5em}
\noindent
{\bf Result} --
Only OctoRPKI and rpstir2 crash when encountering the faulty ROA, whereas all other tested implementations show a warning and move on. 

\subsubsection{ROA ASN overload}

Relying party software has two functions: it retrieves data from repositories, and it processes that information so that RPKI-to-Router (RTR) software can relay that information to a router's BGP Route Origin Validation (ROV) table~\cite{rfc8210}. The memory of a router is limited. If one were to create ROAs for a prefix for every ASN, that means that suddenly the router has to store an extra $2^{32}$ entries. Even at a mere 20~bytes for each entry, this yields around 85~GB of data, which is larger than the memory capacity of most current routers.

\vspace{0.5em}
\noindent
{\bf Result} --
All tested implementations do not limit the maximum amount of ASNs for a prefix. Some implementations do implement some safeguards, such as a maximum repository size or maximum number of files. In all cases, however, we were able to generate a problematic number of entries. The aforementioned safeguards of some RP implementations can likely be circumvented by, for example, delegating part of the ASNs to children, or using rsync exclusively.

\subsubsection{ROA prefix overload}

Much like test J, what can be done with ASNs can also be done with prefixes. If an IPv6 {\tt /48} has been delegated, then one can create $\sum_{i=0}^{80} 2^i \approx 2^{81}$ unique prefixes for one ASN. Combined with the results from J, one can create $2^{81} \times 2^{32} \approx 2^{133}$ entries. Even at 1~bit per entry, this results in the order of $10^{30}$~GB of data. The repository does not need to store this amount of data, as the data can be generated deterministically on the fly, only the SHA-256 hashes need to be stored at 32~bytes per file.

\vspace{0.5em}
\noindent
{\bf Result} --
All tested implementations do not limit the maximum number of subprefixes for a prefix; rpki-client exits with ``excessive runtime (3600 seconds), giving up''. That error message comes up after around four to five hours, rather than the expected one hour. We are unsure what the cause of this is. 

We are not aware of any router that limits the subprefixes it receives over RTR, nor are we aware of any RTR server software that applies its own filtering or aggregation by default. 

\subsubsection{Large file}

This is a simpler version of test D and G. Instead of using elaborate tactics to trick relying party software into processing a lot of data, a lot of data is actually served. For this test, we define the URI attribute for the RRDP snapshot as an URL that refers to a file several gigabytes in size, commonly used for bandwidth tests. We expect similar results to D and G, namely memory exhaustion. This file was hosted using an external URI with random bytes as content.

\vspace{0.5em}
\noindent
{\bf Result} --
OctoRPKI, Fort, rpki-client, and rpstir2 crash when encountering the file. Routinator and RPKI-Prover treat it as a broken repository, and retry over rsync.

\subsubsection{XXE on attributes}

This test tries an XML eXternal Entities (XXE) attack on an attribute. This should not be possible, as XXE are not allowed in attributes. However, as XXE attacks do happen~\cite{cve-2015-3451}, we wanted to ensure that the used XML parsers also behave according to the specifications. In this test, we try to extract contents of a file on the filesystem, and pass them to our server, by including an XML external entity in the snapshot URI.

\vspace{0.5em}
\noindent
{\bf Result} --
All tested implementations do nothing, and neither do the libraries that they use. This is not to say that none of the implementations support XXE, but merely that we have not found a way to exfiltrate data from the host system to the server the repository runs on.

\subsubsection{Long paths}

Much like D, G, and L, this test tries to exhaust the memory of the client running the relying party software, this time by using a path for the data that is much longer than can be reasonably expected (in the order of megabytes). Our expectation is that naively trying to parse this path will result in a crash due to memory exhaustion, and that trying to write to this path will cause an error from the operating system.

\vspace{0.5em}
\noindent
{\bf Result} --
Only OctoRPKI and rpstir2 crash when parsing the long path, whereas all other tested implementations show a warning and move on. We are unsure whether this purely affects the RRDP implementation, or whether this would also be possible over rsync.

\subsubsection{Path traversal}

This test contains valid data with a path that attempts to write to a folder up from where it should. This is based on an rsync security advisory from December 21\textsuperscript{st}, 2015~\cite{rsync_security}, where the server could send a file list containing a path with special folders `{\tt ..}' and `{\tt .}', causing the rsync client to write outside of the destination folder. We attempt to do the same using RRDP, by setting the URI to include `{\tt ..}' in the path. Our expectation is that this potentially allows for remote code execution by being able to write files to arbitrary locations.

\vspace{0.5em}
\noindent
{\bf Result} --
Only OctoRPKI, Fort and rpstir2 allow writing outside the intended destination folder. The other tested implementations reject the path. We observed that default installation instructions make it easy to accidentally (or even intentionally) run OctoRPKI, Fort and rpstir2 as the root user. This means that a file like ``{\tt rsync://example.org/repo/../../etc/cron.daily/ evil.roa}'' could allow for remote code execution on the host machine the relying party software is running on.

\section{Discussion}\label{sec:discussion}
It is important to determine the difficulty of actively executing the attacks discussed in the previous section, as well as their impact in the short term and long term. 

All actively supported implementations had at least one vulnerability that allows them to be disrupted without the need for CA control. This means that an attacker who successfully takes over a repository server, performs a DNS hijack, or has the means to perform a Man-In-The-Middle (MITM) attack, can disrupt the RPKI. As RFC~8182 states that HTTPS certificates may be self-signed~\cite{rfc8182}, this means that any party in the path of any RPKI repository can disrupt RPKI for a large part of the world, without necessarily being part of the RPKI tree. Additionally, controlling part of the RPKI tree is not difficult. In the case of most RIRs, it consists of a small fee and an identity check~\cite{ripe_member} to register as a Local Internet Registry (LIR) with its own resource allocation for which the registrant controls RPKI data and can request delegated RPKI.

Parents, such as RIRs and NIRs, can monitor the behaviour of delegated CAs, and revoke a certificate when malicious behaviour is discovered. The main problem with such an approach is that there is no guarantee that a repository serves the same content to all parties. A malicious party can purposefully exclude the parent from their attack, or direct their attack at only one victim based on the IP address or ASN of the relying party client. An attacker may also only deploy this attack once. This slows down the process of getting the certificate of the malicious party revoked, especially if the malicious party also has non-malicious objects, such as a hijacked legitimate publication point. This is not merely a technical issue, but also requires policy development.

Once the threat has been removed from the tree, the operators of relying party software often also need to manually restart their software. This requires adequate monitoring. From a previous bug in Fort, where the software would crash when encountering a BGPsec certificate~\cite{fort_65}, we observed that roughly 3 out of 10 instances did not come back online within a month. Without adequate monitoring, we would expect the long term effects of our attacks to be similar, meaning that 30\% of relying party instances would not come back online for months after a successful attack, even if that attack only lasted mere hours.

Most of the vulnerabilities we identified have now been resolved, after a coordinated vulnerability disclosure to relying party software implementers that ended with public disclosure on November~9\textsuperscript{th}, 2021 (discussed in Section~\ref{sec:vulnerability-disclosure})\footnote{References to CVEs withheld for double-blind review.}.
%
% TODO for camera ready: unmask these references
% \cite{cve-2021-3907, cve-2021-3908, cve-2021-3909, cve-2021-3910, cve-2021-3911, cve-2021-3912, cve-2021-43172, cve-2021-43173, cve-2021-43174, ncsc-2021-0987}. 
%
However, as of January~1\textsuperscript{st}, 2022, insecure versions of relying party software are still present in the default repositories of many operating system distributions. Ongoing data provided by Kristoff et al.~\cite{kristoff_2020} shows that the now unsupported RIPE NCC RPKI Validator 3 is still used by approximately 5\% of users. Debian~11 still has rpki-client~6.8, OctoRPKI~1.2.2, and Fort~1.5.0 in its default software repository and we found no evidence that the security fixes have been backported. Ubuntu~20.04 still has Fort~1.2.0 in its default software repository. Ubuntu~21.04 and 21.10 also provide rpki-client~6.8 and OctoRPKI~1.2.2. The Fort versions for Ubuntu~21.04 and 21.10 are newer (1.5.0 and 1.5.1 respectively), but are also both still vulnerable. rpstir2 and rcynic are also both still vulnerable with no known fix as of January~1\textsuperscript{st}, 2022.
%TODO Add references to Debian DBAS

\begin{figure}[t]
	\includegraphics[width=0.6\columnwidth]{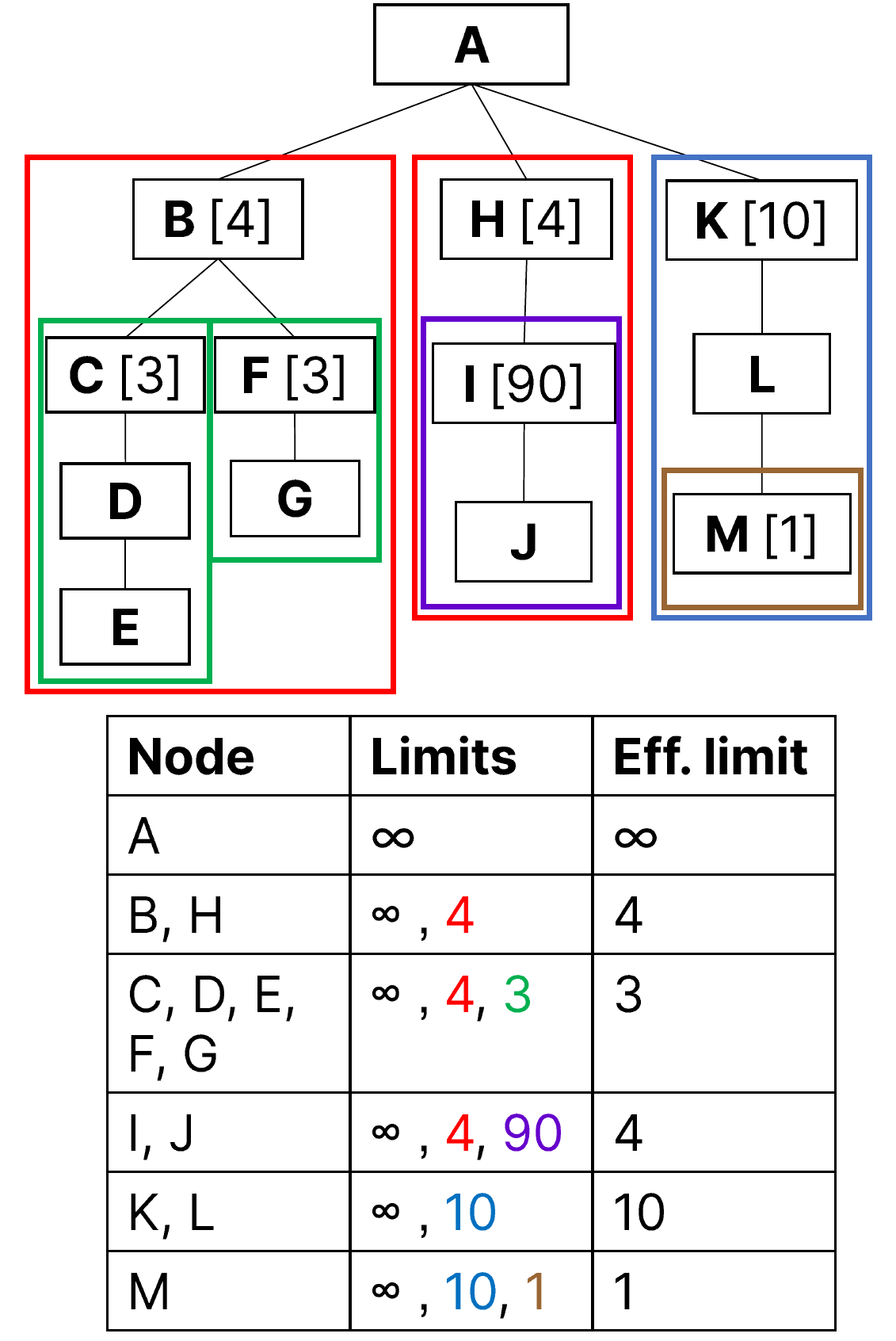}
	\centering
	\caption{An example of how RPKI tree hints could work. Shown is an example tree, with the maximum descendants allowed for each node indicated as a number in brackets. The limit is applied recursively for all descendants, and the effective limit is the strictest of all parents. Since B has an effective limit of 4, and C to G are 5 nodes, relying parties may decide to not retrieve either E or G, depending on the traversal order. What traversal order to use is not specified.}
	\label{fig:tree-hints}
\end{figure}

Additionally, we want to draw the attention to the fact that all implementations are still vulnerable to test H (exponential repository expansion, see Section~\ref{sec:testh}). This is a problem that is thus-far unsolved, as in discussions with relying party software developers we came to the conclusion that there are --- to our knowledge --- no heuristics with which malicious parties can be adequately detected without causing collateral damage to non-malicious parties. When all information of the tree is required, and the nodes in the tree are untrusted, then limits to the structure are required for the tree to be evaluable within a set time limit. The problem is that without outside information, it is not possible to tell which node belongs to the malicious set, and which one does not. Once a parent creates a child, and hands the key to a third party, that third party has exactly the same ability under its subtree as the parent has. Whilst for test A a depth limit is an adequate solution, exponentially increasing repositories already become problematic even at a very shallow depth. Some RIRs and NIRs may have in the order of 10,000~CAs, which makes width restrictions difficult, as our malicious node may be at the same level as NIRs, which may have genuine reasons for their structure. This problem is still present to this day. Setting limits that fit the current shape of the trees, whilst also preventing malicious nodes from being able to disrupt the RPKI, is to our knowledge not possible. This means that either the structure of the RPKI tree must change, certain aspects must be hard-coded in the relying party software, or the parent must be able to provide the required outside information about what can be expected from the child. We have submitted a draft to the IETF SIDROPS working group to add the latter~\cite{kwvanhove-sidrops-rpki-tree-hints-01}. If adopted, this draft would allow a parent to provide ``hints'' about what limits should apply to their children, an example of which is shown in Figure~\ref{fig:tree-hints}.

% !TEX root = ../paper.tex

\section{Ethical considerations}\label{sec:vulnerability-disclosure}
% If the submission deals with vulnerabilities (e.g., software vulnerabilities in a given program or design weaknesses in a hardware system), the authors need to discuss in detail the steps they have already taken or plan to take to address these vulnerabilities (e.g., by disclosing vulnerabilities to the vendors). The same applies if the submission deals with personally identifiable information (PII) or other kinds of sensitive data. If a paper raises significant ethical and legal concerns, it might be rejected based on these concerns.

We realised from the start of our study that we might trigger exploitable vulnerabilities. For this reason, we  decided to test in a manner that would allow us to verify the exploit without disrupting actual day-to-day RPKI operations. Primarily, we decided not to test on the ``production'' RPKI hierarchy, but rather to create our own purpose-built tree hierarchy (as described in Section~\ref{sec:testbed}). 

Over the course of the study, it became clear that we had identified several exploitable security issues. To address these, we started a multi-party coordinated vulnerability disclosure (CVD) process with the help of the National Cyber Security Centre of the Netherlands (NCSC-NL), where the NCSC-NL coordinated the CVD. We decided to enlist the help of NCSC-NL as relying party implementers are from all over the world and coordinating such a multi-party process requires specific expertise. The NCSC-NL is an international frontrunner in developing the CVD process~\cite{ncsc-cvd}. 

The NCSC-NL decided to only contact implementers of actively maintained relying party software, and used the following selection criteria:

\begin{itemize}
	\item The implementer is open to confidential multi-party coordinated vulnerability disclosure~\cite{ncsc-cvd};
	\item The implementer has published clear points of contact to report security issues to.
\end{itemize}

\begin{figure}[t]
	\centering
	\includegraphics[width=\linewidth]{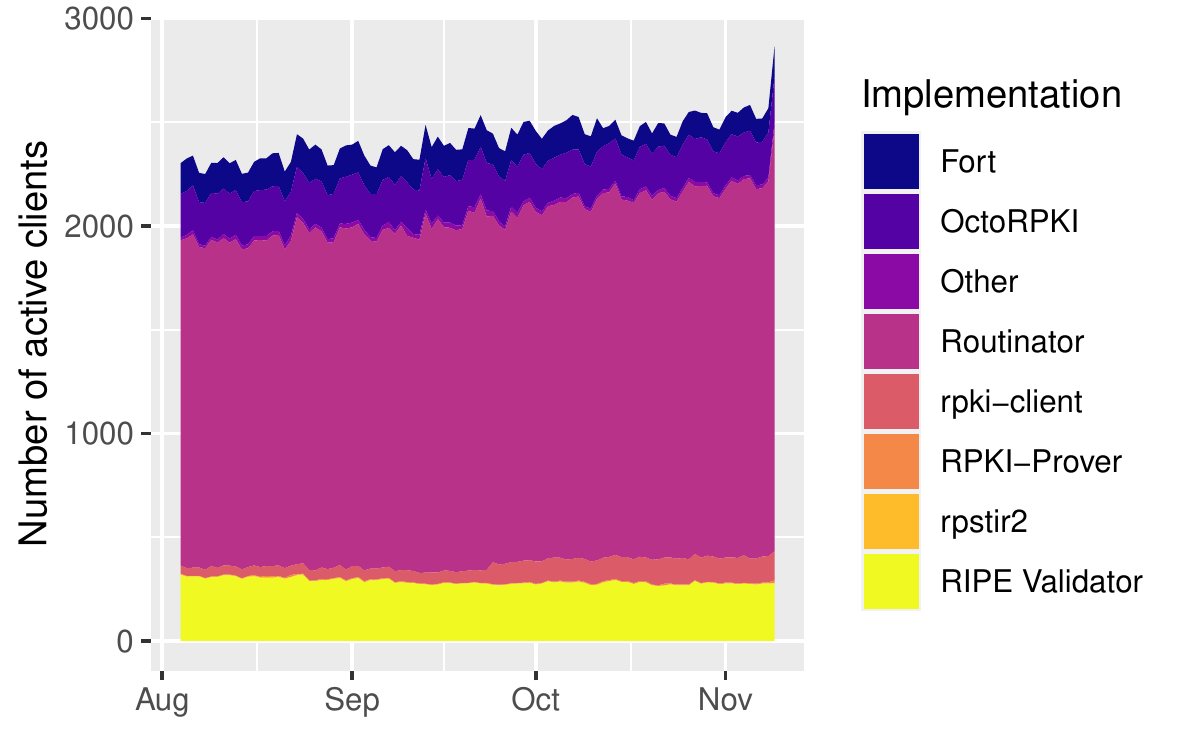}
	\caption{Absolute relying party usage as observed by the RIPE NCC in the months leading up to the disclosure date. All dates are in 2021. Instances of rpki-client on the NLNOG ring are excluded from the count, as these instances are installed by the developers of rpki-client for automated integration testing.}
	\label{fig:rp-graph}
\end{figure}

At the time of starting the CVD process, the parties included were 
\begin{enumerate*}[label=\arabic*)]
	\item Routinator by NLnet Labs,
	\item RPKI Validator 3 by the RIPE NCC, 
	\item OctoRPKI by Cloudflare, and
	\item Fort Validator by NIC M\'{e}xico
\end{enumerate*}. 
NCSC-NL explicitly decided not to contact the developers of rpki-client, from the OpenBSD project, due to their publicly announced policy of full disclosure of vulnerabilities~\cite{openbsd_security} and due to earlier issues with disclosures such as KRACK~\cite{krackattacks}. We note that this decision sparked controversy with the OpenBSD developers, which was eventually resolved in discussions between NCSC-NL and the OpenBSD development team. To enable the OpenBSD team to also resolve the identified vulnerabilities prior to public disclosure, the publication date was postponed by NCSC-NL. For further information we refer to the full timeline of the CVD included in Appendix~\ref{sec:timeline}. 
%
% RvRD: the sentence below lacks context, I suggest we leave it out
%The secondary consideration was ensuring that the users of the various relying party software would not be harmed. 
%
Over the course of the CVD process, we used data provided as a courtesy by the RIPE NCC, shown in Figure~\ref{fig:rp-graph}, to determine the most widely used relying party software. We did not notice any significant change in these figures over the course of the CVD. We also note that at the start of the process (August~4\textsuperscript{th}, 2021) the four RP implementations that were contacted by NCSC-NL together represented 97.5\% of active RP clients according to the data from the RIPE NCC. On the date of disclosure (November~9\textsuperscript{th}, 2021), these four implementations together with rpki-client were used by 98.5\% of active RP clients according to the data from the RIPE NCC. 

Other projects that were not contacted at the start of the CVD process, because they did not meet the criteria set by NCSC-NL, were:

\begin{description}
	\item[rcynic] Developed by Dragon Research Labs, due to the fact that there is no security policy or contact, the software sees little to no active use and the software had not been actively maintained since 2016 as of the time the CVD process started;
	\item[rpstir2] Developed by ZDNS, due to the fact that there is no security policy or contact and the software did not build or work at the time the CVD process started;
	\item[RPKI-Prover] Developed by an individual developer, due to the fact that there is no security policy or contact, and the software saw little or no active production use at the start of the CVD process.
\end{description}

We note that all developers of relying party software (also those not included in the CVD process by NCSC-NL) were notified one week prior to public disclosure, and were given the opportunity to respond and request an extension of the publication deadline. The developer of RPKI-Prover implemented fixes, but did not request a deadline extension, the developers of rpstir2 did not respond to any messages and the developers of rcynic also did not request an extension nor indicate they were intending to implement patches. At the time of submission of this paper, neither rpstir2 nor rcynic have updated their software to protect against the issues we identified and disclosed to them.

%A full and detailed timeline of the CVD can be found in Appendix~\ref{sec:timeline}.
% We have contacted all implementers of relying party software as listed in Table~\ref{fig:test-table} before publication.

% !TEX root = ../paper.tex

\section{Conclusion}\label{sec:conclusion}

In this work, we set out to study to what extent a dishonest RPKI certificate authority or repository publication point could disrupt the operation, and specifically the availability, of relying parties. To do this, we created a threat model and analysed the RPKI specifications and underlying communication protocols. Based on this analysis, we identified fifteen potential attack vectors, which we implemented in a test bench that can be used to evaluate if a relying party software implementation is vulnerable to these attacks.

Extensive testing with eight implementations of relying party software showed that all implementations were vulnerable to at least one of the identified attack vectors. We engaged in a coordinated vulnerability disclosure process with the help of the National Cyber Security Centre of the Netherlands (NCSC-NL) to mitigate these vulnerabilities. We note, though, that some vulnerabilities cannot be trivially resolved by changes to RP software, and for these cases, we have proposed an extension to the RPKI specifications that we are actively engaged in discussing with the community.

%We have looked into the unique characteristics of the RPKI, where unlike other common protocols all information must be retrieved first, and used that to create a threat model where an attacker that wanted to disrupt RPKI operations had full control over a certificate authority and its RRDP publication point. We have developed an exploit framework based on that threat model, and tested how current RPKI relying party software deals with these threats. We showed that in many cases the case where the certificate authority or publication point was malicious was inadequately considered by relying party software, allowing a publication point to disrupt the entirety of RPKI. Additionally, we showed that the protocol design seems in its current state incapable of technically ensuring all the necessary assumptions about the protocol to function properly and securely to hold, making it impossible for relying party software developers to adequately prevent disruption by a malicious certificate authority without collateral damage. Lastly, we have described the steps and considerations for reporting these issues to all parties involved.

\section*{Future work}

We note that there is a general lack of security analysis for the RPKI, as is the case for many other internet protocols. Given the critical nature of the RPKI for the stability of the internet's routing substrate, we believe it necessary to perform a thorough analysis of other aspects of the RPKI, including proposed future extensions that would allow (partial) path validation~\cite{ietf-sidrops-aspa-verification-07}.

Another aspect we intend to study in more detail is the update cycle of relying party software. A preliminary analysis we performed after the public disclosure of the vulnerabilities we identified in this work shows that many operators take a long time to update their RP instances, and some have yet to do so at the time of writing of this paper. Given the security-critical role an RP instance has in an operator's routing decisions, timely updates are of key importance. In this context, it is critical not only to understand the speed at which updates are deployed, but to also study why operators decide to wait.

%So far we have only looked into the RRDP threat model from a relying party perspective. We believe other aspects of the RPKI would benefit from similar scrutiny.

\section*{Acknowledgements}

\emph{Redacted for double-blind review.}
%We would like to thank, in no particular order, the people at: NLnet Labs, RIPE NCC, the National Cyber Security Centre of the Netherlands (NCSC-NL), NIC M\'{e}xico, Cloudflare, and LACNIC for their support during the vulnerability disclosure process.

\section*{Code availability}

\emph{Redacted for double-blind review, we will release the test bed code under a permissive open source licence if the paper is accepted.}
%The code for the test framework, as well as the manual to set it up, is available under the Affero General Public License version 3 on \url{https://gitlab.com/Koenvh/relying-party-resiliency-platform}.

\bibliographystyle{plainurl}
\bibliography{paper}

% !TEX root = ../paper.tex

\clearpage
\appendix
\section{Vulnerability disclosure timeline}\label{sec:timeline}
\newcommand{\foo}{\makebox[0pt]{---}\hskip-0.5pt\vrule width 1pt\hspace{\labelsep}}
Below is the full timeline from the start of the project to the end of the CVD process. All dates are in 2021, times mentioned are in UTC.

\footnotesize
\begin{tabular}{@{\,}r <{\hskip 2pt} !{\foo} >{\raggedright\arraybackslash}p{4cm}}
	%\toprule
	\addlinespace[1.5ex]
	May & Start of the research project\\
	June & First set of vulnerabilities discovered, decision to start a CVD process made, NCSC-NL is contacted by the researchers.\\
	July & Discussion between the researchers and NCSC-NL on the CVD process to follow.\\
	August & Identification of stakeholders in the CVD process to contact.\\
	12 August & Start CVD process -- NCSC-NL contacts relying party software implementers to notify them that vulnerabilities have been found and asks them to confirm willingness to participate under embargo on providing patches with a coordinated release tentatively scheduled for November 8, 2021.\\
	12 August & RIPE NCC responds that they no longer support Validator 3, and that they will not update it.\\
	17 August & The implementers have all responded to the notification and have received information on the vulnerabilities.\\
	25 October & Routinator, OctoRPKI and Fort have patches ready and releases on standby.\\
	25 October & Due to early availability of fixes, publication date is moved forward to November 1, 2021, with notification of all other parties (rpki-client, rcynic, rpstir2, and RPKI-Prover) set for October 27, 2021.\\
	26 October, 16:00h UTC & Pull request with fixes to OctoRPKI is inadvertently made publicly visible. Automated GitHub notifications are sent to all users following this project, including parties not notified yet in the CVD process.\\
	26 October, 18:00h UTC & OctoRPKI pull request is removed from the public record.\\
	26 October, 19:00h UTC & Decision is made to move notification of other parties forward by a few hours in light of the public PR.\\
\end{tabular}

\begin{tabular}{@{\,}r <{\hskip 2pt} !{\foo} >{\raggedright\arraybackslash}p{4cm}}
	%\toprule
	%\addlinespace[1.5ex]
	26 October, 19:00h UTC & Notifications are sent to rpki-client, rcynic, rpstir2, and RPKI-Prover. These notifications are sent by the researcher instead of NCSC-NL, in deviation from the process, in the interest of speed in notification.\\
	27 October & OpenBSD responds to notification of rpki-client, indicating they do not agree to the terms because the time between notification and publication is too short, and not wanting to agree to an embargo \emph{a priori}.\\
	27 October & NCSC-NL gives OpenBSD the option to agree on an entirely new deadline provided OpenBSD agrees to keep the disclosed information confidential under embargo.\\
	27 October & OpenBSD responds negatively to NCSC-NL and indicates they are not willing to agree to an embargo and request not to be contacted again.\\
	27 October & Email to one of the rcynic developers has bounced, and another contact attempt with a corrected email address is made to both developers.\\
	27 October & OpenBSD developers publicly criticise the CVD process in multiple online forums.\\
	29 October & rcynic point of contact publicly denounces CVD process on NANOG mailing list including an unredacted copy of the notification mail in their post.\\
	29 October & NCSC-NL privately notifies the rcynic point of contact about the omission in the notification message describing the rest of the process and timeline.\\
	29 October & NCSC-NL issues a public statement about the CVD process due to ongoing discussion in public forums \cite{ncsc_2021}. \\
	30 October & OpenBSD changes its position and agrees to keep the disclosed information confidential under embargo and is informed of the vulnerabilities.\\
	31 October & On request by the OpenBSD developers the disclosure date is moved to 9 November. \\
	9 November, 14:00h UTC & Embargo is lifted, updates released, CVD process ends.
\end{tabular}

\end{document}